\documentstyle[aps,preprint,psfig,epsf]{revtex}

\input epsf

\tightenlines

\begin{document}

\title{\Large{The Quark-Gluon-Plasma Liquid}}

\author{Markus H. Thoma}

\address{Centre for Interdisciplinary Plasma Science, 
Max-Planck-Institut f\"ur extraterrestrische Physik,
P.O. Box 1312, 85741 Garching, Germany}

\maketitle

\vspace{0.4in}

\begin{abstract}
The quark-gluon plasma close to the critical temperature is a strongly 
interacting system. Using strongly coupled, classical, 
non-relativistic plasmas as an analogy, we argue that the quark-gluon plasma
is in the liquid phase. This allows to understand experimental
observations in ultrarelativistic heavy-ion collisions and to interpret
lattice QCD results. It also supports the indications of the presence of
a strongly coupled QGP in ultrarelativistic heavy-ion collisions.
\end{abstract}

\vspace{0.2in}

Ultrarelativistic heavy-ion collision experiments at the accelerators
SPS (CERN) and RHIC (Brookhaven) are performed to search for a new
state of matter, the so-called quark-gluon plasma (QGP), in which the
quarks and gluons are deconfined \cite{Mueller}. The early Universe
should have been in this phase for the first few microseconds after the
Big Bang. In high-energy
nucleus-nucleus collisions a hot and dense fireball is created which is able to
reach temperatures above the critical one for the transition from the 
hadronic to the QGP phase of the order of $T_c=170$ MeV \cite{Karsch1}.
First indications for the discovery of the QGP in these experiments have 
been reported \cite{Gyulassy,Shuryak,Heinz}. However, the QGP cannot be observed directly, 
as the tiny fireball (diameter about 10 fm) expands and cools down rapidly, 
allowing the QGP phase to exist only for less than about 10 fm/c. Only
hadrons, leptons, and photons emitted from this little bang 
are registered in the detectors. Hence, only by comparing experimental
data with theoretical predictions for signatures of the QGP formation 
the QGP can be detected by circumstantial evidence. Unfortunately,
the theoretical description of the QGP from first principles
is extremely difficult as the interaction between
quarks and gluons in the QGP, described by QCD, is strong. Therefore, 
perturbative QCD is not applicable. Only at extremely high temperatures
the interaction becomes weak due to asymptotic freedom, allowing a 
perturbative treatment of the QGP \cite{Thoma1}. The only so far
well-established non-perturbative method, lattice QCD, is not
capable so far to compute dynamical quantities, such as the most QGP signatures\footnote{An 
exception is the dilepton production rate using the maximum entropy method
\cite{Karsch}.}. 

Consequently, one has to adopt appropriate models or analogies from
other fields in physics \cite{Shuryak} to describe the QGP.
Here we propose to consider strongly coupled, classical, non-relativistic
plasmas as model systems to learn about qualitative features of the 
strongly coupled QGP\footnote{It is quite natural to compare the QGP with
usual plasmas as it shows many features known from classical electrodynamic
plasmas such as plasmons or Debye screening.}.  
Such plasmas are well studied experimentally as well 
as theoretically \cite{Ichimaru}. A non-relativistic plasma is called
strongly coupled if the interaction energy (Coulomb energy) between the 
particles is larger than the thermal energy of the plasma particles, i.e., if
the Coulomb coupling parameter $\Gamma = q^2/(dT)> 1$, where $q$ is the 
charge of the particles, $d$ the interparticle distance, and $T$ the plasma 
temperature ($\hbar=c=k_B=1$). A strongly coupled plasma exhibits peculiar 
properties, in particular liquid and even solid or crystal phases 
\cite{Ichimaru}. For example, in the case of a one-component plasma with
unscreened Coulomb interaction, Monte Carlo simulations of the free energy
revealed a phase transition from the liquid to the solid phase (Coulomb crystal)
if the critical value $\Gamma_c =172$ is exceeded \cite{Slattery}.

We define the Coulomb coupling parameter of the QGP in analogy as $\Gamma = Cg^2/(dT)$
\cite{Thoma2}, where $C=4/3$ or 3 is the Casimir invariant for the quarks or
gluons, respectively, and $g$ the strong coupling constant, related to the strong
fine structure constant by $\alpha_s=g^2/(4\pi)$. For an estimate of the 
parton-parton interaction strength we have replaced here the 
square of the electric charge $e^2$ in the Coulomb potential
by the square of the strong charge $Cg^2$. Note, however, that the heavy
quark potential, for example in the case of a charmonium, is given by
$V(r)=C\alpha_s/r+{\rm confining}$ potential, i.e. the Coulomb part differs by a 
factor of $4\pi$ from our parton interaction energy entering $\Gamma$. 
For light quarks and gluons, however,
we consider the scattering amplitude for estimating the interaction strength,
where the electric charge $e$ in QED is replaced by $g$ times a color factor 
in QCD \cite{Cutler}.  

For typical temperatures\footnote{Here we consider only an equilibrated QGP. 
However, in ultrarelativistic heavy-ion collisions the fireball is not in 
equilibrium at the beginning. Starting from a color glass condensate in the high 
energy nucleons, the fireball is expected to end up in an equilibrated
QGP phase via a pre-equilibrium stage. How this transformation from
the weakly interacting color glass condensate to a strongly coupled QGP
proceeds by initial hard and secondary parton collisions, including parton
production by bremsstrahlung and pair creation, is the subject of intense 
and on-going investigations \cite{Gelis}.} that can
be reached in heavy-ion collisions, $\Gamma $ is larger than 1. For example,
for $T=200$ MeV the temperature dependent strong fine structure constant
$\alpha_s$ is between 0.2 and 0.5, corresponding to $g=1.5$ - 2.5.
For simplicity, we assume for the particle density the ideal gas result (for
2 active quark flavors in the QGP)
$n=(34/\pi^2)\> \zeta (3)\> T^3$, leading to $d\simeq n^{-1/3} \simeq 0.5$ fm.
This gives $\Gamma = 10$ - 30. Note that we have neglected the chromomagnetic
interaction here, which is only possible in the non-relativistic case. Hence, 
the parton interaction is larger and the QGP even stronger coupled. 
Plasmas with such a value of $\Gamma$ are known to be in the liquid phase, 
i.e., the plasma behaves rather like a liquid than a gas
due to the strong interaction between the charged particles. Therefore we 
suggest that the QGP, at least close to the critical temperature,
is also in the liquid phase. This was also suggested by Gyulassy and Heinz 
\cite{Gyulassy,Heinz}
considering the success of hydrodynamical models for describing RHIC data.

Relativistic effects do not alter this picture
as they even increase the interaction energy. What about quantum effects?
In a plasma quantum effects become important if the thermal de Broglie
wave length $\lambda_{th}$ exceeds the interparticle distance. In
the QGP we estimate this wave length by $\lambda_{th}=1/m^*$, where
$m^*$ is the effective parton mass in the QGP. For gluons we take the 
effective mass from perturbative QCD, $m^*=gT/\sqrt{3}$ (for quarks,
$m^*=gT/\sqrt{6}$), which is a good estimate even at $T$ not too far
from $T_c$ \cite{Peshier}. Then we find $\lambda_{th} \simeq 1$ fm,
which is of the same order as $d=0.5$ fm, showing that quantum effects 
are important. However, they do not destroy
the assumption of a QGP liquid because of the following argument: 
quantum effects are dominant in degenerate
plasmas, such as quark matter. In this case the thermal energy in the 
denominator of the Coulomb coupling parameter $\Gamma$ has to be replaced by 
the Fermi energy ($E_F \simeq 350$ MeV), reducing $\Gamma $ by less than
a factor 1/2 compared to a QGP at $T=200$ MeV. 
In a degenerate electron plasma Wigner crystallization sets in if
$\Gamma \simeq 15$ \cite{Ichimaru}, indicating that also quark matter is
in the liquid or even solid phase. 
Of course, non-abelian contributions to the 
parton interaction will be important for quantitative considerations.
Hence we conclude that for a qualitative understanding of the QGP at
temperatures, reached in heavy-ion collisions, the comparison with
strongly coupled, classical, non-relativistic plasmas can be useful.

However, strongly coupled plasmas are difficult to produce in the laboratory 
and rarely exist in nature. The reason is simple: a strongly coupled
plasma requires a high density corresponding to a short interparticle
distance and/or a low temperature. Under these conditions the ions and 
electrons recombine quickly and the plasma vanishes\footnote{Even the plasma
in the center of the sun is weakly coupled, although the density is 130 g/cm$^3$, because
of the high temperature $T=1.5 \times 10^7$ K.}. Exceptions are, for
example,
the ion component in a white dwarf ($\Gamma = 10$ - 200) \cite{Ichimaru}
or dense, 
short-living plasmas created by heavy-ion beams on solid state targets
\cite{Dewald}. Another possibility are so-called complex or dusty plasmas 
\cite{Bouchoule,Merlino}. These are multi-component low-temperature plasmas, created 
by external electromagnetic fields (rf or dc fields), in which, in addition
to ions, electrons, and neutral atoms, microparticles, i.e., micron
size particles like dust, are present. These microparticles get 
charged by collecting electrons and ions on their surface. Due to the high mobility of 
the electrons compared to the ions the net charge of the particles can be of the
order of $10^3$ to $10^5$ electron charges, depending on the size of the 
particles and the electron temperature. Owing to this high charge, the 
microparticles interact strongly by the Yukawa force with each other,
corresponding to a large $\Gamma\gg 1$. As a matter of fact, the formation
of a plasma crystal in complex plasmas has been predicted \cite{Ikezi}
and experimentally verified \cite{Thomas1,Thomas2} as shown in Fig.1. Another advantage
of complex plasmas is their easy observability. Using an illumination laser and a CCD camera the
microparticles can be observed directly, providing informations of the
system on the microscopic and kinetic level.

Now let us discuss two possible consequences of our assumption that the QGP 
is in the liquid phase.

1. A liquid shows a pronounced
peak in the pair correlation function\footnote{In the case of a crystal, 
on the other hand, the pair correlation function 
consists of a number of sharp peaks corresponding to the distance to the nearest 
neighbor, to the next to the nearest neighbor, and so on. Also in a liquid 2 or 3 peaks
can be observed in the pair correlation function or the structure factor, related by Fourier 
transformation. However, only 
the first one corresponding to the nearest neighbors is significant, whereas the following 
ones are small and broad \cite{Ichimaru,Hansen}.}, defined in coordinate space by
\cite{Thomas2,Hansen}
\begin{equation}
g(r)=\left \langle \frac{1}{N}\> \sum_{i\neq j}^N \delta ({\bf r}-{\bf r_i}-
{\bf r_j}) \right \rangle ,
\label{Pair}
\end{equation}
where ${\bf r_i}$ and ${\bf r_j}$ are the position vectors of 2 particles 
and $N$ the total particle number. An example for the pair correlation
function of the liquid phase of a complex plasma is shown in fig.2. 

In QCD lattice simulations of the temporal meson correlators a strong correlation of the 
propagating $q\bar q$-pair above $T_c$ was found \cite{Karsch2}. These features of the meson 
correlators cannot be explained in a weakly coupled QGP \cite{Karsch3} but have been interpreted instead 
as the existence of bound states in the QGP \cite{Shuryak}. Note, however, that there is a qualitative 
change if $T_c$ is exceeded: the sharp $\delta$-like peak below $T_c$ in the spectral function
of the correlators, corresponding to mesonic bound states, turns into a broad ``resonance'' above 
$T_c$ \cite{Karsch2}. Therefore, we suggest that the meson correlators above $T_c$ rather indicate 
a liquid QGP than a bound state in the QGP. To prove this assumption QCD lattice calculations of 
static parton pair correlators would be helpful. 

2. Strong interactions in a plasma imply that the standard Coulomb 
scattering theory is not valid because standard Coulomb scattering theory 
works only if the Coulomb radius, $\rho=q^2/E$, of a particle with energy $E$
is much smaller than the Debye 
length $\lambda_D=1/\mu$. In the QGP close to $T_c$ we use the non-perturbative
estimate $\mu=6T$ \cite{Laine} for the Debye screening mass corresponding to a Debye screening length
of less than 0.2 fm at $T=200$ MeV. The Coulomb radius for a thermal parton with energy $3T$
is given by $\rho=Cg^2/3T=1$ - 6 fm. Therefore we get $\beta = \rho/\lambda_D =5$ - 30.
Consequently the Debye screening length cannot be used as cutoff for 
calculating the transport cross section \cite{Landau}
from the differential t-channel cross section. This means that in  
a strongly coupled plasma the interaction range is much
larger than the Debye screening length. This has been considered in the case
of the interaction of the ions with the microparticles in a
complex plasma \cite{Khrapak}, where ions with an impact parameter
larger than the Debye screening length have been shown to be important.
As a result a modified Coulomb logarithm $\Lambda^*$
was derived by considering all ions that approach the microparticle closer than 
$\lambda_D$, yielding a cutoff for the impact parameter which can be 
several times the Debye screening length.
This leads to a transport cross section ($\sigma_t \sim \Lambda^*$), which can be more
than an order of magnitude larger than predicted by the standard Coulomb
scattering theory. Whereas the standard Coulomb logarithm in the transport
cross section is given by $\Lambda =(1/2)\ln (1+1/\beta^2)$, the modified 
reads $\Lambda^*=\ln (1+1/\beta)$, which is valid for about $\beta <10$
\cite{Khrapak}. For instance, for $\beta =10$ $\Lambda = 0.005$ and $\Lambda^*=0.1$
follows, leading to an enhancement of the transport cross section by a factor 
of 20. 

This could explain the enhancement of the parton transport cross sections, which was  
concluded from the elliptic flow data \cite{Molnar}, and the fast 
thermalization of the fireball
in relativistic heavy-ion collisions which is required to explain the 
success of the hydrodynamical description of hadron spectra \cite{Heinz},
even if perturbative results are used. In order to estimate this effect in the QGP, 
we consider the transport cross section for 
elastic gluon scattering given in Eq. (3) of Ref.\cite{Molnar}:
\begin{equation}
\sigma_t (s) = 4\> \sigma_0\> z\> (1+z)\> [(2z+1)\> \ln (1+1/z)- 2],
\label{e2}
\end{equation}  
where  $\sigma_0 = 9\pi \alpha_s^2/(2\mu^2)$ and $z=\mu^2/s$ with the Debye mass $\mu $ and
the square of the center of mass scattering energy $s$. The Debye screening
mass $\mu$ was used as an IR cutoff to obtain finite results. In a strongly coupled plasma
with $\beta =10$, the maximum impact parameter is given by $4.6 \lambda_D$ (see Eq. (9) in
Ref.\cite{Khrapak}). Assuming now this value also for the QGP, we replace $\mu$ in 
(\ref{e2})
by $\mu /4.6\simeq 1.3 T$. Assuming, in addition, thermal energies for the gluons, leading
to $s\simeq 15 T^2$ \cite{Thoma3}, we find
a cross section enhancement by a factor of 13.    

For higher parton energies $E$ the Coulomb radius $\rho=Cg^2/E$ becomes smaller
and the transport cross section reduces to the perturbative results \cite{Thoma3}.
Of course, their might be additional sources for a cross section enhancement, such as 
higher order and non-perturbative contributions to the scattering amplitudes. Furthermore,
non-linear screening effects in strongly coupled plasmas can lead to an interparticle 
potential different from a Yukawa potential,
which results in a reduced screening and therefore enhanced scattering cross sections.
Non-linear short range modifications of the screening potential are known, for instance, 
to play an important role in the enhancement
of thermonuclear reaction rates in the late stages of stellar evolution \cite{Ichimaru}.

Summarizing, strongly coupled plasmas have peculiar properties, 
in particular liquid or solid phases. Comparing with strongly-coupled, classical
non-relativistic plasmas, such as complex plasmas, we conclude that the QGP
is in the liquid phase. This may help to discover the QGP rather than 
looking for high-temperature limit
(ideal QGP) properties from  perturbative QCD or static properties from 
lattice QCD.
Indeed, the QGP liquid explains qualitatively observed features of
relativistic heavy-ion collisions, such as flow, cross section enhancement, 
and fast thermalization, and can be used to interpret lattice QCD results,
e.g., meson correlators. Presumably, the examples discussed above are only
a small selection for comparing properties of the QGP with liquid systems.
We expect that there will be a number of interesting features of the data observed
in ultrarelativistic heavy-ion collisions which can be understood by the analogy
to liquid plasmas. For example, the equation of state and the quark number susceptibilities
should be analyzed in the liquid phase, although the deviation from the weak coupling
regime is not so pronounced in this case \cite{Blaizot1,Mustafa}.

For a more quantitative description, we propose to use molecular dynamics simulations, similar 
as in the case of strongly coupled plasmas, but now for a classical, relativistic Yukawa system
consisting of partons with an effective, temperature dependent mass. 
In this way, quantitative results
for the equation of state, correlation functions, and other interesting 
quantities may be derived. Thus lessons and methods from the physics of strongly coupled plasmas
may open up a new field of ``high-energy plasma physics'' or ``plasma-particle physics''
besides the transport description (Vlasov equation) of weakly coupled plasmas used for
relativistic plasmas (electron-positron plasma, high-temperature QGP) already since the sixties
\cite{Silin,Blaizot2}.  

The analogy to complex plasmas supports the evidences found so far in ultrarelativistic
heavy-ion collisions for the discovery of a strongly coupled QGP. Such a system is even more
interesting than an ideal or perturbative QGP, testing a new phase of non-perturbative QCD besides
the confinement phase.

\bigskip

{\bf Acknowledgments}

\medskip

The author would like to thank U. de Angelis, S. Datta, U. Heinz, E. Laermann, and M. Laine for 
valuable discussions and encouragement.

\begin{figure}
\epsfxsize=4.5in
\epsfbox{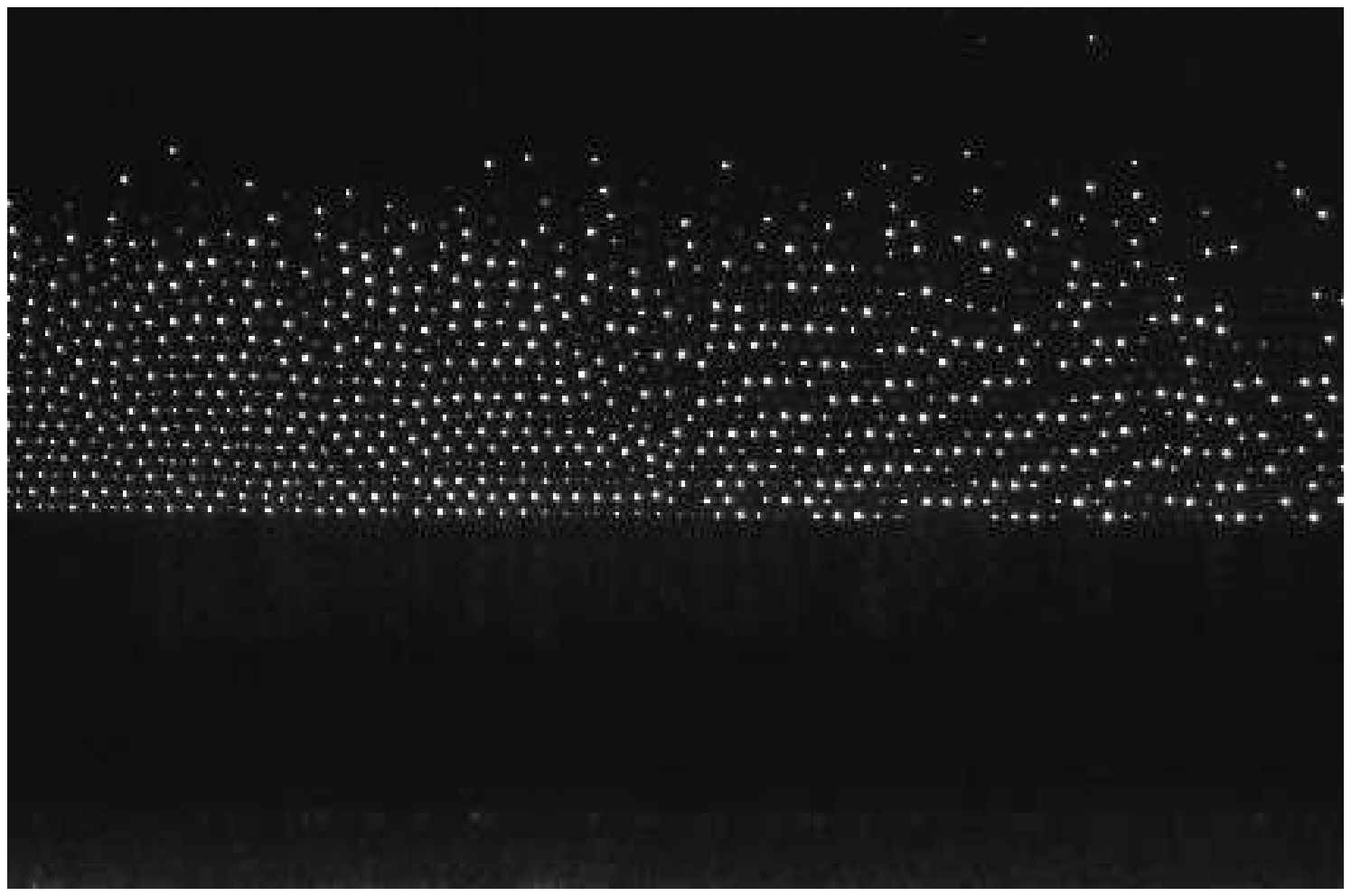}
\vspace{0.2in}

\noindent{Fig.1: Hexagonal structures in a complex plasma (plasma crystal). 
The microparticles of radius 3.4 microns
are illuminated by a laser and recorded by a CCD camera. The interparticle distance is about
200 microns.}
\end{figure}

\begin{figure}
\epsfxsize=4.5in
\epsfbox{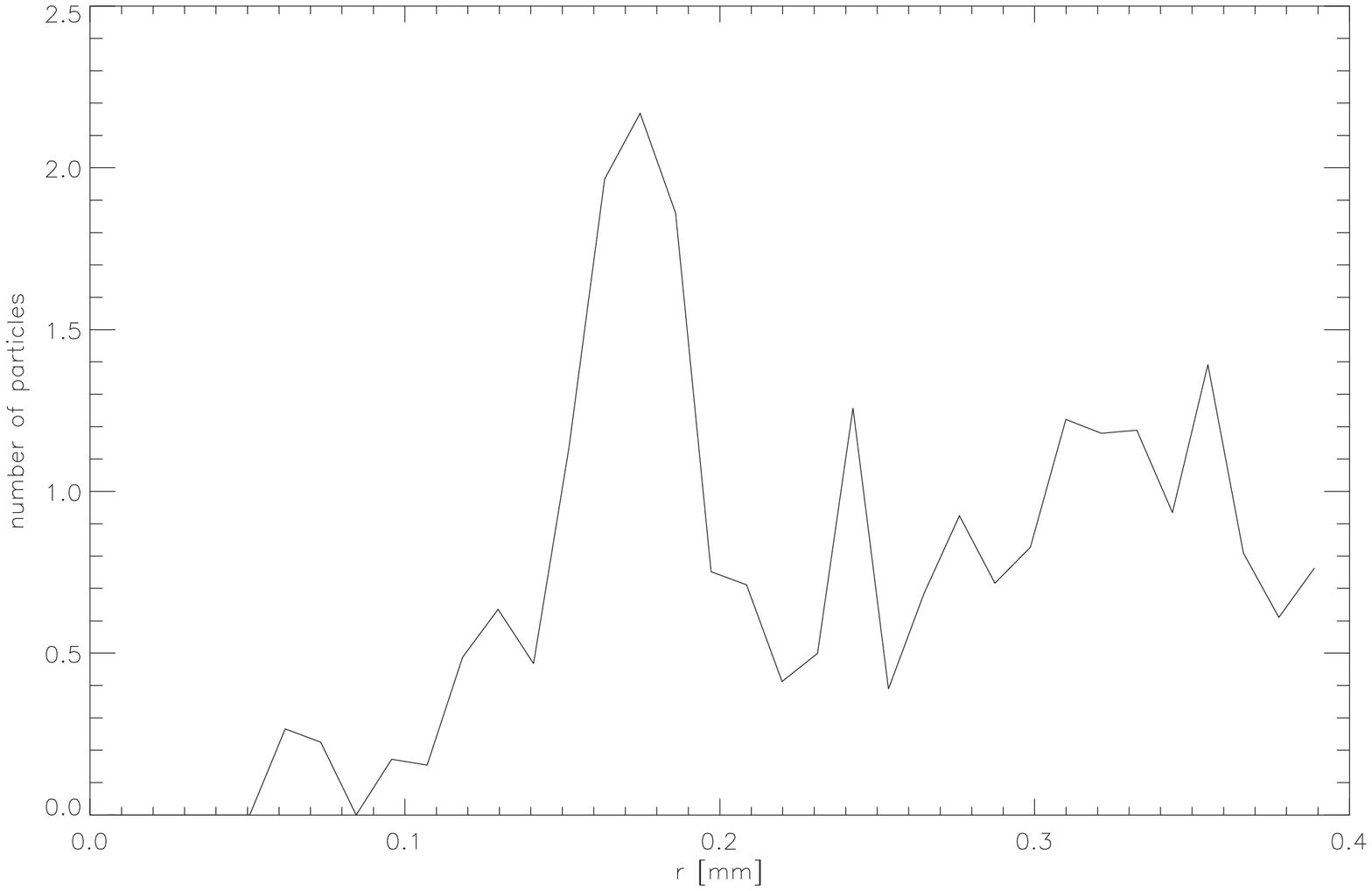}
\vspace{0.2in}

\noindent{Fig.2: Pair correlation function in the liquid phase of a complex plasma. The pronounced
peak at $r=0.2$ mm indicates the interparticle spacing. The other structures are statistical
fluctuations.}
\end{figure}


\end{document}